\RequirePackage{fix-cm}
\documentclass[5p,sort&compress]{elsarticle}
\usepackage{textcomp}
\usepackage[utf8]{inputenc} 
\usepackage{amsmath,amssymb,graphicx,tabularx,textcomp,gensymb,lscape}
\usepackage{siunitx}\usepackage{microtype}
\DeclareSIUnit{\atomicpercent}{at.\%}
\DeclareSIUnit{\wtpercent}{wt.\%}
\DeclareSIUnit{\angstrom}{\text{\AA}}
\DeclareSIUnit{\h}{\hour}
\usepackage{float}  
\usepackage{etoolbox}
\makeatletter
\renewcommand\normalsize{\@setfontsize\normalsize{9}{11}}
\makeatother
\AtBeginDocument{\normalsize}
\usepackage{xcolor}
\usepackage[normalem]{ulem}
\definecolor{DeleteRed}{RGB}{200,0,0}
\definecolor{AddGreen}{RGB}{0,128,0}

\begin{document}

\begin{frontmatter}
\title{A New High $\gamma$--$\gamma'$ Fraction Polycrystalline Co/Ni Superalloy}

\author[IC]{Cameron Crabb*}\ead{c.crabb23@ic.ac.uk}
\author[RR]{Mark C. Hardy}
\author[IC]{Michele S. Conroy}
\author[IC]{David Dye*}\ead{david.dye@imperial.ac.uk}

\address[IC]{Department of Materials, Royal School of Mines, Imperial College London, Prince Consort Road, London, SW7 2BP, UK}
\address[RR]{Rolls-Royce plc., Elton Road, Derby, DE24 8BJ, UK}

\begin{abstract}
\noindent An ingot-processable Co/Ni-based $\gamma/\gamma'$ superalloy has been developed for wrought parts in aircraft engines. An \SI{8.5}{\kilogram} VIM ingot was homogenised and hot rolled to a true strain of 1. A heat treatment was then developed to produce \SI{50}{\percent} $\gamma'$ and a grain size of \SI{142}{\micro\meter}. High-temperature tensile testing showed yield strengths of \SI{820}{\mega\pascal} at \SI{700}{\degreeCelsius}, \SI{785}{\mega\pascal} at \SI{800}{\degreeCelsius} and \SI{783}{\mega\pascal} at \SI{850}{\degreeCelsius}. Creep tests at \SI{750}{\degreeCelsius}/\SI{600}{\mega\pascal} and \SI{800}{\degreeCelsius}/\SI{300}{\mega\pascal} demonstrated reduced minimum creep rates and extended rupture lives relative to current wrought Ni-base alloys. Oxidation at \SI{800}{\degreeCelsius} for \SI{100}{\hour} formed a \SI{1}{\micro\metre} dual chromia/alumina scale with minimal subsurface degradation. Long-term exposure at \SI{800}{\degreeCelsius} for \SI{8000}{\hour} showed limited $\gamma'$ coarsening and no deleterious intermetallic or grain boundary phases. The alloy therefore offers a promising balance of processability, strength and oxidation resistance for high-temperature static and rotating components.
\end{abstract}

\end{frontmatter}

High-temperature static aerospace engine components such as turbine casings must sustain load for extended periods and resist oxidation whilst remaining readily manufacturable \textcolor{black}{\cite{Reed2006TheApplications,Pollock2006}}. Many wrought Ni-base alloys operate near practical temperature limits in these roles, which subsequently motivates the development of alloys that can retain strength and creep resistance at higher temperature whilst maintaining environmental resistance \textcolor{black}{\cite{Neumeier2015,Neumeier2024}}. These parts are commonly large and axisymmetric, and are well suited to ring roll forging. Ring rolling offers efficient production of large-diameter sections with reduced material waste relative to extensive machining from large forgings or billets \textcolor{black}{\cite{Allwood2005,Stanistreet2006}}. This manufacturing route imposes strict constraints on workability, defect tolerance, and process consistency across large sections.

Co/Ni-based $\gamma$--$\gamma'$ alloys occupy a useful compositional space for such applications because high $\gamma'$ fractions support yield strength and creep resistance, while a sufficiently low $\gamma'$ solvus can provide a wide temperature window for hot working in the $\gamma$ region, and suitable Cr/Al levels can support the formation of protective oxide scales during long exposures \textcolor{black}{\cite{Neumeier2015,Knop2014ASuperalloy,Ismail2017Oxidation}}. At the same time, these objectives are coupled. Increasing $\gamma'$ fraction and solid-solution strengthening generally raises flow stress and can increase susceptibility to cracking during large-scale deformation \textcolor{black}{\cite{Neumeier2024}}. Changes made to improve oxidation resistance can also alter phase stability and grain-boundary chemistry, with consequences for processing and long-term performance. Heat-treatment-controlled grain-boundary morphology can also influence workability; cooling-induced serration improved the hot-compression formability of U720Li \textcolor{black}{\cite{Chen2023HotWorkability}}. Early V208-family compositions demonstrated promising strength and creep performance, motivating further optimisation of creep resistance, microstructural stability and ingot processability \textcolor{black}{\cite{Knop2014ASuperalloy,McAuliffe2021,Vorontsov2022}}.

Alloy design for ring-rolled static components is governed by trade-offs between strengthening, oxidation behaviour, and environmentally assisted intergranular cracking, whilst avoiding segregation-driven cracking, formation of brittle phases, and grain-boundary damage during prolonged exposure \textcolor{black}{\cite{McAuliffe2021,Kontis2016,Fan2020}}. Al supports $\gamma'$ formation and can contribute to protective scale development, whilst Cr promotes chromia formation and can slow external and internal oxidation at intermediate temperatures \textcolor{black}{\cite{Ismail2017Oxidation,Wo2023Wagner}}. Refractory additions such as Nb, W and Ta can increase creep resistance through solid-solution strengthening and by modifying $\gamma/\gamma'$ partitioning and phase stability; excessive additions can promote unwanted intermetallic phases \textcolor{black}{\cite{Fan2020,Fleischmann2015,Lapington2018}}. Ta, Nb and Ti can modify planar fault energies in $\gamma'$, with effects on precipitate shearing that depend on composition and fault type \textcolor{black}{\cite{Kumar2018}}. Minor elements are commonly used to control grain-boundary chemistry and carbide/boride populations, which can improve boundary stability and damage tolerance, but can also degrade boundary performance if phase fraction, morphology, or distribution are not well controlled \textcolor{black}{\cite{McAuliffe2021,Kontis2016,Yan2014AlloyingAlloys}}.

The alloy reported here is derived from the V208 family and adjusted to suit ingot-based processing and ring-rolling manufacture whilst targeting higher high-temperature capability \textcolor{black}{\cite{Knop2014ASuperalloy,McAuliffe2021,Hardy2024US11898228B2}}. The chemistry is selected to maintain a high $\gamma'$ fraction, limit segregation sensitivity, and retain oxidation resistance through controlled Cr and Al additions. The alloy is processed through an industrially relevant thermomechanical route and evaluated in terms of microstructural stability, yield strength, creep response under relevant conditions, and oxidation resistance with associated environmental sensitivity at elevated temperature.
\begin{table*}[!tb]
\centering
\caption{Nominal alloy compositions in at.\%}
\label{tab:alloy_compositions}
\renewcommand{\arraystretch}{1.0}
\setlength{\tabcolsep}{4pt}
\normalsize
\begin{tabular}{lcccccccccccccc}
\hline
        & \textbf{Ni} & \textbf{Co} & \textbf{Cr} & \textbf{W} & \textbf{Mo} &
          \textbf{Al} & \textbf{Ta} & \textbf{Nb} & \textbf{Ti} &
          \textbf{Si} & \textbf{Hf} & \textbf{C} & \textbf{B} & \textbf{Zr} \\
\hline
Alloy 5 \cite{Hardy2024US11898228B2}  & 51.60 & 20.00 & 12.00 & 2.00 & 1.75 & 9.75 & 1.00 & 0.50 & 1.25 & --   & --   & 0.075 & 0.05 & 0.02 \\
RR1000 \cite{Christofidou2018}                                  & 50.95 & 17.90 & 16.50 & --   & 3.00 & 6.35 & 0.60 & --   & 4.30 & --   & 0.16 & 0.13  & 0.08 & 0.03 \\

Waspaloy \cite{Neumeier2015}                                & 56.40 & 13.00 & 21.30 & --   & 2.60 & 2.70 & --   & --   & 3.60 & --   & --   & 0.38  & 0.03 & --   \\
U720Li \cite{Neumeier2015}                                 & 54.90 & 14.30 & 17.30 & 0.40 & 1.80 & 5.20 & --   & --   & 5.90 & --   & --   & 0.12  & 0.09 & 0.02 \\

\hline
\end{tabular}
\end{table*}

An ingot was vacuum induction melted (VIM) at the Royce Facility in Sheffield prior to a multi-stage homogenisation treatment in air (\SI{1160}{\degreeCelsius} for \SI{6}{\hour} + \SI{1210}{\degreeCelsius} for \SI{66}{\hour}) based on Thermo-Calc predictions, and then hot rolled at \SI{1220}{\degreeCelsius} to a total true strain $=1$, also at the Royce Facility. The resulting \SI{8.5}{\kilogram} rolled plate was heat treated at Imperial College to achieve an optimal $\gamma'$ fraction, distribution and morphology using first a \SI{30}{\minute} solution treatment at \SI{1160}{\degreeCelsius}, followed by cooling at \SI{0.3}{\degreeCelsius\per\minute} through the $\gamma'$ solvus to \SI{1060}{\degreeCelsius} where it was held for \SI{30}{\minute}. The alloy was then cooled at \SI{60}{\degreeCelsius\per\minute} to \SI{850}{\degreeCelsius} and aged for \SI{3}{\hour}. It was subsequently cooled at \SI{0.3}{\degreeCelsius\per\minute} to \SI{800}{\degreeCelsius}, held for \SI{2}{\hour}, and finally furnace cooled to room temperature. DSC was performed using a Mettler Toledo DSC/TGA~2 to obtain an accurate $\gamma'$ solvus value and workability range of the alloy. Microstructural analysis was performed using a Tescan Clara scanning electron microscope (SEM); meanwhile TEM for STEM-EDX was performed on a Thermo-Fisher Spectra operating at \SI{300}{\kilo\volt}. Creep tests were conducted in tension at \SI{750}{\degreeCelsius} under a \SI{600}{\mega\pascal} load and at \SI{800}{\degreeCelsius} under a \SI{300}{\mega\pascal} load. Slow strain-rate tensile testing was carried out at \SI{800}{\degreeCelsius} in vacuum and air at a strain rate of \SI{1e-4}{\per\second} on a Zwick-Roell testing frame fitted with a \SI{5}{\kilo\newton} load cell, and all other tensile tests performed at a strain rate of \SI{1e-3}{\per\second} from room temperature up to temperatures of \SI{850}{\degreeCelsius}. Oxidation tests were performed on specimens measuring \SI{10}{\milli\metre}~$\times$~\SI{10}{\milli\metre}~$\times$~\SI{2}{\milli\metre}, ground and polished by standard metallographic procedure, and oxidised at \SI{800}{\degreeCelsius} for \SI{100}{\hour} in static air. Long term stability was examined after exposure at \SI{800}{\degreeCelsius} for \SI{8000}{\hour} in air.

The fully heat-treated microstructure of the investigated alloy is shown in \textbf{Figure ~1a-c} in addition to the long term stability shown in \textbf{Figure ~1d}. The alloy shows a uniform grain distribution with an average grain size of \SI{142}{\micro\metre}. This is important for an ingot-route alloy of this chemistry: as-cast microsegregation in heavily alloyed systems can produce interdendritic regions susceptible to incipient melting during overly aggressive homogenisation \textcolor{black}{\cite{Jablonski2009,Pang2016SolutionTreatment}}. BSE imaging of the finalised microstructure showing homogeneous grain distribution. In-column backscattered SEM imaging, \textbf{Figure ~ 1c}, reveals a $\gamma/\gamma'$ microstructure characterised by grain boundary primary $\gamma'$, relatively large ogdoadically diced secondary intragranular $\gamma'$ precipitates and a finer dispersion of tertiary $\gamma'$, defined by its size being under 50nm. The secondary $\gamma'$ phase exhibits a faceted, segmented appearance, where individual cubes appear subdivided into smaller units by internal contrast, giving rise to the ogdoadically diced morphology. Split cuboidal $\gamma'$ morphologies, including eight-cuboid configurations, have previously been documented in Ni-based alloys \textcolor{black}{\cite{Doi1984Shape}}. The tertiary $\gamma'$ is more uniform and spherical and occupies the $\gamma$ channels between the larger precipitates, contributing to the overall high $\gamma'$ fraction. The development of distinct secondary and tertiary $\gamma'$ populations is consistent with the sensitivity of precipitate size distributions to cooling history reported in Ni-based superalloys \textcolor{black}{\cite{Chen2015Cooling}}. A total $\gamma'$ fraction of 50\% is achieved with a tertiary fraction of approximately 10\%. The $\gamma'$ fraction was calculated using the Blavette method from atom probe tomography needles \textcolor{black}{\cite{Blavette1988,Blavette2000TheSuperalloys}}.

\begin{figure*}[!htbp]
\centering
 \includegraphics[width=0.98\textwidth]{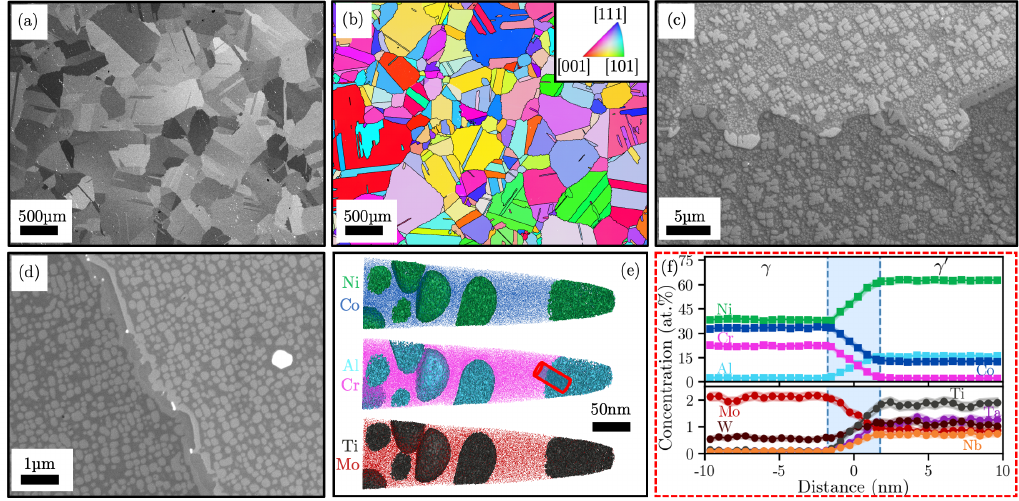}
  \caption{SEM and EBSD characterisation of the heat-treated Alloy~5 microstructure. 
(a) Backscattered-electron SEM image showing the recrystallised grain structure, 
(b) EBSD inverse pole figure map, 
(c) in-lens SEM image showing the $\gamma/\gamma^\prime$ precipitate distribution, (d) In-Lens Backscatter image after \mbox{\SI{8000}{\hour} }\hskip0pt showing the long term stability of the alloy. (e) and (f) APT needles and 1D concentration profile showing the phase composition and partitioning of analysed ions}
  \label{fig:Micro}
\end{figure*}     

DSC results show a $\gamma'$ solvus of \SI{1110}{\degreeCelsius} with solidus temperature of \SI{1250}{\degreeCelsius}. This results in a wide processing window for thermomechanical processing such as ring rolling: the alloy can be homogenised and worked in air at temperatures that are high enough to reduce flow stresses and promote dynamic recovery, and below the measured bulk solidus. Local incipient-melting temperatures in segregated regions must nevertheless be considered when selecting a homogenisation schedule. Local melting of $\mathrm{M}_3\mathrm{B}_2$ borides has also been directly linked to intergranular cracking during hot compression of a Ni-based superalloy with elevated B and C contents \textcolor{black}{\cite{Schulz2023IncipientMelting}}. At the same time, the $\gamma'$ solvus is positioned such that a high retained $\gamma'$ fraction is present at the service relevant temperatures, rather than collapsing the ordered phase fraction during operation. At the same time, the alloy is stable against intermetallic phase formation after long exposures, such as $\sigma$, shown in \textbf{Figure ~1d}. 

APT was used to analyse the phase composition and partitioning behaviour, with typical Ni/Co-base partitioning displayed, \textbf{Figure ~1e-f}. As expected $\gamma'$ formers such as Al, Ni, Ti and Nb partition strongly, while $\gamma$ formers in Ni-base systems such as Cr, Co and Mo partition accordingly. However, there is still a relatively large amount of Co within the $\gamma'$ phase, in addition to slight partitioning of W to the $\gamma'$ phase. This highlights a phenomenon seen in Ni/Co-based systems where the $(\mathrm{Co},\mathrm{Ni})_3(\mathrm{Al},\mathrm{W})~\gamma^\prime$ phase is precipitated \textcolor{black}{\cite{Knop2014ASuperalloy,Shinagawa2008}}. W additions can modify $\gamma'$ coarsening through changes in phase chemistry and diffusion kinetics; however, W partitioning alone does not establish the rate-controlling species in the present alloy \textcolor{black}{\cite{Baler2020}}. In addition, no deleterious phases such as $\sigma$ are observed at the grain boundaries or within the grain interiors.

Grain boundaries are decorated with intergranular secondary $\gamma'$ and exhibit pronounced serration, as also reported in slowly cooled V208-derived Co/Ni alloys \textcolor{black}{\cite{McAuliffe2021}}. In Ni-based superalloys, serrated grain-boundaries have been associated with improved creep rupture lives and improved resistance to time-dependent intergranular crack growth, although these depend on testing conditions \textcolor{black}{\cite{Hong2009Serration,Tang2018}}. The $\gamma'$ volume fraction is higher than in most wrought Ni-based superalloys, and comparable to high-$\gamma'$ wrought alloys, while retaining the hot-rolling capability \textcolor{black}{\cite{Reed2006TheApplications,Neumeier2024}}. Compared with conventional casing alloys, the fraction of ordered phase is significantly higher, moving the design closer to disc-like chemistries. The density of the alloy (\SI{8.39}{\gram\per\centi\metre\cubed}) lies slightly above that of typical Ni-base casing alloys but below many Co-rich $\gamma/\gamma'$ alloys, so the penalty in specific strength is limited. Taken together, the microstructural observations and phase stability measurements suggest that the alloy occupies a useful intermediate space between powder Co/Ni superalloys and traditional wrought Ni-base alloys.

\begin{table}[!ht]
\centering
\caption{APT-derived phase compositions and partitioning behaviour between the $\gamma$ and $\gamma'$ phases in the heat-treated alloy.}
\label{tab:apt_partitioning}
\renewcommand{\arraystretch}{1}
\setlength{\tabcolsep}{5pt}
\normalsize
\begin{tabular}{lcccc}
\hline
\textbf{Element} &
\textbf{$C_{\gamma}$} &
\textbf{$C_{\gamma'}$} &
\textbf{$K$} &
\textbf{Partitioning} \\
\hline
Ni & 38.0 & 63.0 & 1.6  & $\gamma'$ \\
Al & 2.4  & 16.0 & 6.5  & $\gamma'$ \\
Ti & 0.1  & 1.9  & 18   & $\gamma'$ \\
Ta & 0.1  & 1.3  & 14   & $\gamma'$ \\
W  & 0.6  & 1.1  & 1.9  & $\gamma'$ \\
Nb & 0.1  & 0.7  & 9.2  & $\gamma'$ \\
Co & 33.0 & 12.5 & 0.4  & $\gamma$ \\
Cr & 22.0 & 2.2  & 0.1  & $\gamma$ \\
Mo & 2.1  & 0.9  & 0.41 & $\gamma$ \\
\hline
\end{tabular}
\end{table}

Tensile results shown in \textbf{Figure ~2} show that, despite the larger grain size, the presented alloy achieves yield strengths comparable or exceeding commercial wrought Ni-base superalloys across the temperature range investigated \cite{SpecialMetals2004Waspaloy,Heck2020Waspaloy}. The alloy exhibits a room-temperature 0.2\% proof stress of \SI{920}{\mega\pascal}, decreasing to \SI{820}{\mega\pascal} at \SI{700}{\degreeCelsius} and \SI{783}{\mega\pascal} at \SI{850}{\degreeCelsius}. At elevated temperature, the 0.2\% proof stress remains competitive with Waspaloy in its fine-grained state ASTM 6-8, demonstrating that the high $\gamma^\prime$ fraction provides effective strengthening without preventing ingot processing and hot rolling. The retained strength is consistent with the high $\gamma^\prime$ volume fraction, refractory-element partitioning and the serrated grain-boundary microstructure developed during heat treatment.

Slow strain rate tensile testing at \SI{800}{\degreeCelsius}, shown in \textbf{Figure~2}, was used to assess susceptibility to environmentally assisted intergranular cracking, following the approach of Gabb et al. \textcolor{black}{\cite{Gabb2014SSRT}}. Comparative tests performed in vacuum and in air assess the effect of oxidation-assisted damage. The reduction in ductility in air indicates some environmental sensitivity at \SI{800}{\degreeCelsius}. These monotonic tests do not directly measure fatigue crack-growth resistance. Grain-boundary $\gamma^\prime$ and boundary chemistry can influence oxygen-assisted damage, but a protective mechanism cannot be inferred from the tensile response alone \textcolor{black}{\cite{Nemeth2017Environment}}.

Creep testing shown in \textbf{Figure ~3} demonstrates that Alloy~5 provides better creep resistance than the comparison Ni-base alloys in its current configuration \cite{Christofidou2018,Gu2009,Gabb2002ME3,SpecialMetals2004Waspaloy,Vorontsov2022}. The measured minimum creep rates were approximately \SI{1.2e-8}{\per\second} and \SI{4.4e-10}{\per\second}, respectively. At all conditions, the alloy exhibits good minimum creep rates and longer rupture lives than RR1000, Udimet~720Li and Waspaloy under comparable testing conditions. This improved creep response is attributed to the combined effect of the high $\gamma^\prime$ fraction, large grain size, grain-boundary serration and refractory-element additions. The high $\gamma^\prime$ fraction provides resistance to dislocation motion within the grains, while serrated grain boundaries are expected to reduce grain-boundary sliding and delay intergranular damage. Refractory additions such as Mo may also contribute through matrix solid-solution strengthening \textcolor{black}{\cite{Fleischmann2015}}. Overall, the mechanical data show that Alloy~5 combines competitive tensile strength with improved creep resistance and acceptable environmental tolerance at \SI{800}{\degreeCelsius}, supporting its potential use in static aerospace components.
\begin{figure}[t]
\centering
 \includegraphics[width=\columnwidth]{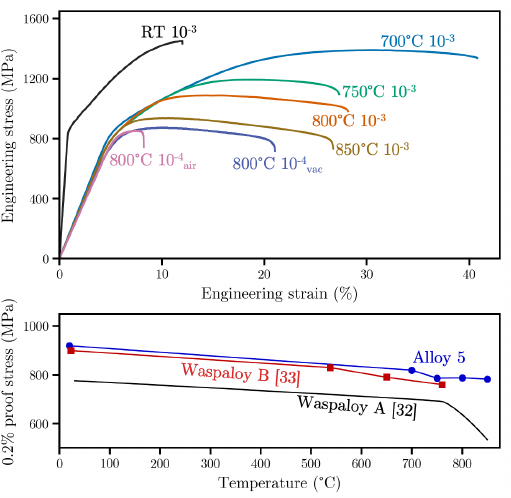}
\caption{Alloy~5 stress--strain curves at \SI{1e-3}{\per\second}, including \SI{800}{\degreeCelsius} air/vacuum tests at \SI{1e-4}{\per\second}, and 0.2\% proof stress compared with Waspaloy datasets A and B.}
\label{fig:Creep2}
\end{figure}
\begin{figure}[t]
\centering
\includegraphics[width=1\columnwidth]{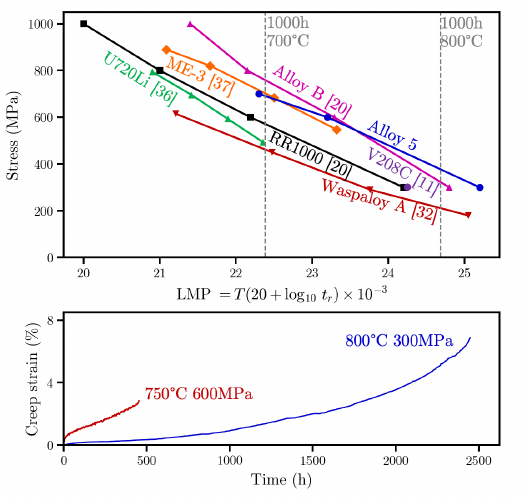}
\caption{Larson--Miller comparison of Alloy~5 creep rupture behaviour against Ni-base reference alloys and associated creep curves for reported tests.}
\label{fig:Creep}
\end{figure}

The oxidation response of Alloy~5 after exposure at \SI{800}{\degreeCelsius} for \SI{100}{\hour} is summarised in \textbf{Figure ~4}. Cross-sectional STEM--EDX shows a thin external oxide scale, approximately \SI{1}{\micro\metre} thick, with a Cr-rich outer region and Al enrichment towards the metal/oxide interface. Ti and Ta, which partition to $\gamma^\prime$ in the unoxidised alloy, are also locally enriched within the scale. Their presence indicates redistribution of these additions during oxidation, although elemental enrichment alone does not establish the formation of separate oxide phases. This layered chemistry is consistent with the chromia/alumina-based oxidation response reported for a related wrought Co/Ni-based alloy, for which the time-dependent scale evolution and elemental redistribution have been examined in detail \textcolor{black}{\cite{Crabb2026Oxidation}}.

The formation of a thin Cr/Al-rich oxide scale is an important result for a Co/Ni-based $\gamma/\gamma^\prime$ alloy. Early Co-rich $\gamma/\gamma^\prime$ systems were often limited by rapid growth of thick, non-protective cobalt oxides, which restricted their environmental capability despite promising high-temperature strength \textcolor{black}{\cite{Klein2011HighSuperalloys}}. In the present alloy, the balanced Co/Ni chemistry, together with controlled Cr and Al additions, suppresses the formation of a thick Co-rich external scale and instead promotes a more protective chromia/alumina-containing oxide, consistent with composition-dependent oxidation trends in Co/Ni alloys \textcolor{black}{\cite{Ismail2017Oxidation,Crabb2026Oxidation,Li2022NiCrOxidation}}. The Cr-rich outer scale is consistent with a barrier to rapid oxidation, while Al enrichment at the metal/oxide interface is consistent with an alumina subscale; the continuity and transport properties of that subscale are not established by elemental maps alone \textcolor{black}{\cite{Crabb2026Oxidation,Wo2024TheSuperalloy}}.

\begin{figure}[!ht]
\centering
\includegraphics[width=\columnwidth]{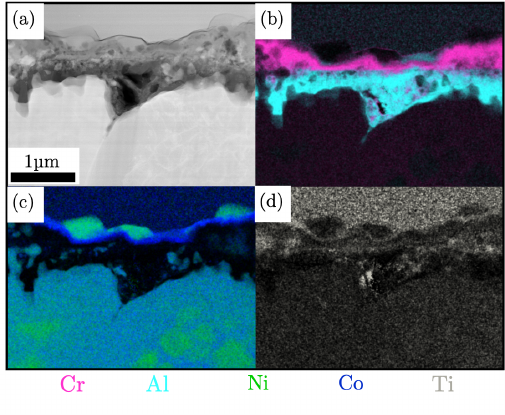}
\caption{Cross-sectional STEM--EDX analysis of Alloy~5 after oxidation at \SI{800}{\degreeCelsius} for \SI{100}{\hour}. (a) HAADF image and (b--d) elemental maps showing the distribution of the principal oxide-forming elements. HAADF from \cite{Crabb2026Oxidation}.}
\label{fig:Oxidation}
\end{figure}

Beneath the oxide, the $\gamma/\gamma^\prime$ microstructure remains largely intact after \SI{100}{\hour} at \SI{800}{\degreeCelsius}. No extensive internal oxidation, deep grain-boundary oxide penetration or large-scale dissolution of the strengthening microstructure is observed over the timescale examined. This is significant because the alloy is designed for static aero-engine components where environmental resistance must be retained alongside creep strength. The limited subsurface degradation observed here indicates that the oxidation response does not rapidly compromise the near-surface $\gamma/\gamma^\prime$ microstructure under the present exposure condition \textcolor{black}{\cite{Crabb2026Oxidation}}.

The oxidation data therefore complement the mechanical results by demonstrating that the alloy retains environmental resistance at a temperature relevant to the targeted application. Together with the tensile and creep behaviour, the formation of a thin chromia/alumina-containing scale at \SI{800}{\degreeCelsius} supports the broader alloy-design objective: an ingot-processable Co/Ni-based $\gamma/\gamma^\prime$ superalloy that combines high-temperature mechanical capability with oxidation resistance suitable for static aero-engine components.

In summary, a novel ingot-processable Co/Ni-based $\gamma$--$\gamma'$-strengthened superalloy has been developed for high-\hspace{0pt}temperature static aero-engine applications. The alloy was produced by VIM, homogenised, hot rolled to a true strain of 1 and heat treated to generate a stable $\gamma/\gamma'$ microstructure containing approximately \qty{50}{\percent} $\gamma'$, including approximately \qty{10}{\percent} tertiary $\gamma'$, segmented secondary $\gamma'$ precipitates and serrated, $\gamma'$-decorated grain boundaries. DSC measurements show a $\gamma'$ solvus of \qty{1110}{\degreeCelsius}, with solidus temperature of \qty{1250}{\degreeCelsius}, providing a wide processing window for ingot processing and ring rolling. The heat-treated alloy exhibits yield strengths comparable to, or higher than, established wrought Ni-base superalloys such as RR1000, Udimet 720Li and Waspaloy, while creep testing at \qty{750}{\degreeCelsius}/\qty{600}{\mega\pascal} and \qty{800}{\degreeCelsius}/\qty{300}{\mega\pascal} demonstrates lower minimum creep rates and longer rupture lives than these reference alloys. Further improvements in yield strength are expected through grain refinement via Hall--Petch strengthening and optimisation of the $\gamma'$ size distribution, particularly the large intragranular secondary $\gamma'$ precipitates. Oxidation at \qty{800}{\degreeCelsius} for \qty{100}{\hour} produces a thin Cr/Al-rich protective scale, approximately \qty{1}{\micro\meter} in thickness, with limited subsurface degradation, and long-term exposure at \qty{800}{\degreeCelsius} for \qty{8000}{\hour} shows limited $\gamma'$ coarsening and no deleterious grain-boundary phases. These results show that Alloy 5 offers a promising balance of processability, microstructural stability, creep strength and oxidation resistance for ingot-processed polycrystalline aero-engine components operating at elevated temperature.
\par\medskip
\textbf{Acknowledgements} \par 
The authors would like to acknowledge support from Rolls-Royce plc for their continued support and funding. MC acknowledges funding from Royal Society Tata University Research Fellowship URF/R1/201318, Royal Society Enhancement Award RF/ERE/210200EM1, and ERC CoG DISCO grant 101171966.

\medskip

{\color{black}
\bibliographystyle{scripta-numbered-1a}\bibliography{references}

@book{Reed2006TheApplications,
    title = {{The Superalloys: Fundamentals and Applications}},
    year = {2006},
    author = {Reed, Roger C.},
    publisher = {Cambridge University Press},
    isbn = {9780521859042},
  doi = {10.1017/CBO9780511541285}
}

@article{Pollock2006,
  author = {Pollock, Tresa M. and Tin, Sammy},
  title = {{Nickel-Based Superalloys for Advanced Turbine Engines: Chemistry, Microstructure, and Properties}},
  journal = {Journal of Propulsion and Power},
  year = {2006}, volume = {22}, number = {2}, pages = {361--374},
  doi = {10.2514/1.18239},
  shortjournal = {J. Propuls. Power}
}

@article{Neumeier2015,
  author = {Neumeier, S. and Freund, L. P. and G{\"o}ken, M.},
  title = {{Novel wrought {$\gamma$}/{$\gamma^{\prime}$} cobalt base superalloys with high strength and improved oxidation resistance}},
  year = {2015},
  journal = {Scripta Materialia},
  volume = {109},
  pages = {104--107},
  doi = {10.1016/j.scriptamat.2015.07.030},
  shortjournal = {Scr. Mater.}
}

@article{Neumeier2024,
  author = {Neumeier, S. and Freund, L. P. and Bezold, A. and K{\"o}brich, M. and Vollh{\"u}ter, J. and Hausmann, D. and Solis, C. and Stark, A. and Schell, N. and Pyczak, F. and Felfer, P. and Gilles, R. and G{\"o}ken, M.},
  title = {{Advanced Polycrystalline {$\gamma^{\prime}$}-Strengthened CoNiCr-Based Superalloys}},
  year = {2024},
  journal = {Metallurgical and Materials Transactions A},
  volume = {55},
  pages = {1319--1337},
  doi = {10.1007/s11661-024-07319-6},
  shortjournal = {Metall. Mater. Trans. A}
}

@article{Allwood2005,
  author = {Allwood, J. M. and Tekkaya, A. E. and Stanistreet, T. F.},
  title = {{The Development of Ring Rolling Technology}},
  journal = {Steel Research International},
  year = {2005}, volume = {76}, number = {2--3}, pages = {111--120},
  doi = {10.1002/srin.200505981},
  shortjournal = {Steel Res. Int.}
}

@article{Stanistreet2006,
  author = {Stanistreet, T. F. and Allwood, J. M. and Willoughby, A. M.},
  title = {{The design of a flexible model ring rolling machine}},
  journal = {Journal of Materials Processing Technology},
  year = {2006}, volume = {177}, number = {1--3}, pages = {630--633},
  doi = {10.1016/j.jmatprotec.2006.03.228},
  shortjournal = {J. Mater. Process. Technol.}
}

@article{Knop2014ASuperalloy,
    title = {{A New Polycrystalline Co-Ni Superalloy}},
    year = {2014},
    journal = {JOM},
    author = {Knop, M. and Mulvey, P. and Ismail, F. and Radecka, A. and Rahman, K. M. and Lindley, T. C. and Shollock, B. A. and Hardy, M. C. and Moody, M. P. and Martin, T. L. and Bagot, P. A. J. and Dye, D.},
    number = {12},
    pages = {2495--2501},
    volume = {66},
    publisher = {Minerals, Metals and Materials Society},
    doi = {10.1007/s11837-014-1175-9},
    issn = {15431851},
  shortjournal = {JOM}
}

@article{Ismail2017Oxidation,
 author = {Ismail, F. B. and Vorontsov, V. A. and Lindley, T. C. and Hardy, M. C. and Dye, D. and Shollock, B. A.},
 title = {Alloying effects on oxidation mechanisms in polycrystalline {Co--Ni} base superalloys},
 journal = {Corrosion Science},
 volume = {116},
 year = {2017},
 pages = {44--52},
 doi = {10.1016/j.corsci.2016.12.009},
  shortjournal = {Corros. Sci.}
}

@article{McAuliffe2021,
  author = {McAuliffe, T. P. and Bantounas, I. and Reynolds, L. R. and Foden, A. and Hardy, M. C. and Britton, T. B. and Dye, D.},
  title = {{Quantitative Precipitate Classification and Grain Boundary Property Control in Co/Ni-Base Superalloys}},
  year = {2021},
  journal = {Metallurgical and Materials Transactions A},
  volume = {52},
  pages = {1649--1664},
  doi = {10.1007/s11661-021-06156-1},
  shortjournal = {Metall. Mater. Trans. A}
}

@article{Vorontsov2022,
  author = {Vorontsov, V. A. and McAuliffe, T. P. and Hardy, M. C. and Dye, D. and Bantounas, I.},
  title = {{Precipitate dissolution during deformation induced twin thickening in a CoNi-base superalloy subject to creep}},
  year = {2022},
  journal = {Acta Materialia},
  volume = {232},
  pages = {117936},
  doi = {10.1016/j.actamat.2022.117936},
  shortjournal = {Acta Mater.}
}

@article{Kontis2016,
  author = {Kontis, P. and Mohd Yusof, H. A. and Pedrazzini, S. and Danaie, M. and Moore, K. L. and Bagot, P. A. J. and Moody, M. P. and Grovenor, C. R. M. and Reed, R. C.},
  title = {{On the effect of boron on grain boundary character in a new polycrystalline superalloy}},
  year = {2016},
  journal = {Acta Materialia},
  volume = {103},
  pages = {688--699},
  doi = {10.1016/j.actamat.2015.10.006},
  shortjournal = {Acta Mater.}
}

@article{Fan2020,
  author = {Fan, Zhongding and Wang, Xinguang and Yang, Yanhong and Chen, Hao and Yang, Zhigang and Zhang, Chi},
  title = {{Microstructural Instability and Precipitation Behaviors of Intermetallic Phases in a Nb-Containing CoNi-Based Superalloy}},
  year = {2020},
  journal = {Metallurgical and Materials Transactions A},
  volume = {51},
  pages = {2495--2508},
  doi = {10.1007/s11661-020-05710-7},
  shortjournal = {Metall. Mater. Trans. A}
}

@article{Wo2023Wagner,
 author = {Wo, J. W. X. and Hardy, M. C. and Stone, H. J.},
 title = {Evaluating {Wagner} Oxidation Criteria for Protective {$\mathrm{Al}_2\mathrm{O}_3$} Scale Formation in {Ni}-Based Superalloys},
 journal = {High Temperature Corrosion of Materials},
 volume = {100},
 number = {1--2},
 year = {2023},
 pages = {85--108},
 doi = {10.1007/s11085-023-10163-5},
  shortjournal = {High Temp. Corros. Mater.}
}

@article{Fleischmann2015,
  author = {Fleischmann, Ernst and Miller, Michael K. and Affeldt, Ernst and Glatzel, Uwe},
  title = {{Quantitative experimental determination of the solid solution hardening potential of rhenium, tungsten and molybdenum in single-crystal nickel-based superalloys}},
  year = {2015},
  journal = {Acta Materialia},
  volume = {87},
  pages = {350--356},
  doi = {10.1016/j.actamat.2014.12.011},
  shortjournal = {Acta Mater.}
}

@article{Lapington2018,
  author = {Lapington, M. T. and Crudden, D. J. and Reed, R. C. and Moody, M. P. and Bagot, P. A. J.},
  title = {{Characterization of Phase Chemistry and Partitioning in a Family of High-Strength Nickel-Based Superalloys}},
  year = {2018},
  journal = {Metallurgical and Materials Transactions A},
  volume = {49},
  pages = {2302--2310},
  doi = {10.1007/s11661-018-4558-7},
  shortjournal = {Metall. Mater. Trans. A}
}

@article{Kumar2018,
  author = {Kumar, Kaushlendra and Sankarasubramanian, R. and Waghmare, Umesh V.},
  title = {{Tuning planar fault energies of Ni3Al with substitutional alloying: First-principles description for guiding rational alloy design}},
  year = {2018},
  journal = {Scripta Materialia},
  volume = {142},
  pages = {74--78},
  doi = {10.1016/j.scriptamat.2017.08.021},
  shortjournal = {Scr. Mater.}
}

@article{Yan2014AlloyingAlloys,
    title = {{Alloying effects in polycrystalline {$\gamma$}$'$ strengthened Co--Al--W base alloys}},
    year = {2014},
    journal = {Intermetallics},
    author = {Yan, H.-Y. and Vorontsov, V. A. and Dye, D.},
    month = {5},
    pages = {44--53},
    volume = {48},
    doi = {10.1016/j.intermet.2013.10.022},
    issn = {09669795},
  shortjournal = {Intermetallics}
}

@misc{Hardy2024US11898228B2,
  author = {Hardy, Mark C. and Dye, David and Reynolds, Lucy and McAuliffe, Thomas and Bantounas, Ioannis},
  title = {{Alloy}},
  year = {2024},
  howpublished = {U.S. Patent US11898228B2, granted 13 February 2024},
  note = {Assignee: Rolls-Royce plc},
  url = {https://patents.google.com/patent/US11898228B2/en}
}

@article{Jablonski2009,
  author = {Jablonski, Paul D. and Cowen, Christopher J.},
  title = {{Homogenizing a Nickel-Based Superalloy: Thermodynamic and Kinetic Simulation and Experimental Results}},
  year = {2009},
  journal = {Metallurgical and Materials Transactions B},
  volume = {40},
  number = {2},
  pages = {182--186},
  doi = {10.1007/s11663-009-9227-1},
  shortjournal = {Metall. Mater. Trans. B}
}

@article{Pang2016SolutionTreatment,
  author = {Pang, Hon Tong and D'Souza, Neil and Dong, Hongbiao and Stone, Howard J. and Rae, Catherine M. F.},
  title = {Detailed Analysis of the Solution Heat Treatment of a Third-Generation Single-Crystal Nickel-Based Superalloy {CMSX-10K}},
  journal = {Metallurgical and Materials Transactions A},
  year = {2016},
  volume = {47},
  pages = {889--906},
  doi = {10.1007/s11661-015-3252-2},
  shortjournal = {Metall. Mater. Trans. A}
}

@inproceedings{Blavette1988,
  author = {Blavette, D. and Caron, P. and Khan, T.},
  title = {{An Atom-Probe Study of Some Fine-Scale Microstructural Features in Ni-Based Single Crystal Superalloys}},
  year = {1988},
  booktitle = {Superalloys 1988},
  editor = {Reichman, S. and Duhl, D. N. and Maurer, G. and Antolovich, S. and Lund, C.},
  publisher = {The Metallurgical Society},
  pages = {305--314},
  doi = {10.7449/1988/Superalloys_1988_305_314},
}

@article{Blavette2000TheSuperalloys,
    title = {{The Role of the Atom Probe in the Study of Nickel-Based Superalloys}},
    year = {2000},
    journal = {Materials Characterization},
    author = {Blavette, D. and Cadel, E. and Deconihout, B.},
    number = {1-2},
    month = {1},
    pages = {133--157},
    volume = {44},
    doi = {10.1016/S1044-5803(99)00050-9},
    issn = {10445803},
  shortjournal = {Mater. Charact.}
}

@article{Shinagawa2008,
  author = {Shinagawa, Kazuya and Omori, Toshihiro and Sato, Jun and Oikawa, Katsunari and Ohnuma, Ikuo and Kainuma, Ryosuke and Ishida, Kiyohito},
  title = {{Phase Equilibria and Microstructure on {$\gamma'$} Phase in Co-Ni-Al-W System}},
  journal = {Materials Transactions},
  year = {2008}, volume = {49}, number = {6}, pages = {1474--1479},
  doi = {10.2320/matertrans.MER2008073},
  shortjournal = {Mater. Trans.}
}

@article{Baler2020,
  author = {Baler, Nithin and Pandey, Prafull and Palanisamy, Dhanalakshmi and Makineni, Surendra Kumar and Phanikumar, Gandham and Chattopadhyay, Kamanio},
  title = {{On the effect of W addition on microstructural evolution and {$\gamma^{\prime}$} precipitate coarsening in a Co--30Ni--10Al--5Mo--2Ta--2Ti alloy}},
  year = {2020},
  journal = {Materialia},
  volume = {10},
  pages = {100632},
  doi = {10.1016/j.mtla.2020.100632},
  shortjournal = {Materialia}
}

@article{Tang2018,
  author = {Tang, Yuanbo T. and Wilkinson, Angus J. and Reed, Roger C.},
  title = {{Grain Boundary Serration in Nickel-Based Superalloy Inconel 600: Generation and Effects on Mechanical Behavior}},
  year = {2018},
  journal = {Metallurgical and Materials Transactions A},
  volume = {49},
  pages = {4324--4342},
  doi = {10.1007/s11661-018-4671-7},
  shortjournal = {Metall. Mater. Trans. A}
}

@incollection{Gabb2014SSRT,
 author = {Gabb, Tim P. and Telesman, Jack and Banik, Anthony and McDevitt, Erin},
 title = {Use of Slow Strain Rate Tensile Testing to Assess the Ability of Several Superalloys to Resist Environmentally-Assisted Intergranular Cracking},
 booktitle = {8th International Symposium on Superalloy 718 and Derivatives},
 publisher = {John Wiley \& Sons},
 year = {2014},
 pages = {697--712},
 doi = {10.1002/9781119016854.ch54}
}

@article{Nemeth2017Environment,
 author = {N{\'e}meth, A. A. N. and Crudden, D. J. and Armstrong, D. E. J. and Collins, D. M. and Li, K. and Wilkinson, A. J. and Grovenor, C. R. M. and Reed, R. C.},
 title = {Environmentally-assisted grain boundary attack as a mechanism of embrittlement in a nickel-based superalloy},
 journal = {Acta Materialia},
 volume = {126},
 year = {2017},
 pages = {361--371},
 doi = {10.1016/j.actamat.2016.12.039},
  shortjournal = {Acta Mater.}
}

@article{Crabb2026Oxidation,
 author = {Crabb, Cameron and Kloenne, Zachary T. and Rogers, Samuel R. and Kwok, Chi Hang Dominik and Hardy, Mark C. and Conroy, Michele S. and Dye, David},
 title = {Time-Dependent Oxidation and Scale Evolution of a Wrought {Co/Ni}-Based Superalloy},
 journal = {Advanced Engineering Materials},
 year = {2026},
 volume = {28},
 number = {17},
 pages = {e202503059},
 doi = {10.1002/adem.202503059},
  shortjournal = {Adv. Eng. Mater.}
}

@article{Klein2011HighSuperalloys,
 author = {Klein, L. and Bauer, A. and Neumeier, S. and G{\"o}ken, M. and Virtanen, S.},
 title = {High temperature oxidation of {$\gamma/\gamma'$}-strengthened {Co}-base superalloys},
 journal = {Corrosion Science},
 volume = {53},
 number = {5},
 year = {2011},
 pages = {2027--2034},
 doi = {10.1016/j.corsci.2011.02.033},
  shortjournal = {Corros. Sci.}
}

@article{Wo2024TheSuperalloy,
 author = {Wo, J. W. X. and Hardy, M. C. and Stone, H. J.},
 title = {The Effect of {Nb}, {Ta}, and {Ti} on the Oxidation of a New Polycrystalline {Ni}-Based Superalloy},
 journal = {High Temperature Corrosion of Materials},
 volume = {101},
 number = {3},
 year = {2024},
 pages = {485--509},
 doi = {10.1007/s11085-023-10218-7},
  shortjournal = {High Temp. Corros. Mater.}
}

@article{Christofidou2018,
  author = {Christofidou, K. A. and Hardy, M. C. and Li, H.-Y. and Argyrakis, C. and Kitaguchi, H. and Jones, N. G. and Mignanelli, P. M. and Wilson, A. S. and Mess{\'e}, O. M. D. M. and Pickering, E. J. and Gilbert, R. J. and Rae, C. M. F. and Yu, S. and Evans, A. and Child, D. and Bowen, P. and Stone, H. J.},
  title = {{On the Effect of Nb on the Microstructure and Properties of Next Generation Polycrystalline Powder Metallurgy Ni-Based Superalloys}},
  year = {2018},
  journal = {Metallurgical and Materials Transactions A},
  volume = {49},
  pages = {3896--3907},
  doi = {10.1007/s11661-018-4682-4},
  shortjournal = {Metall. Mater. Trans. A}
}

@article{Gu2009,
  author = {Gu, Y. F. and Cui, C. and Ping, D. and Harada, H. and Fukuda, T. and Fujioka, J.},
  title = {{Creep behavior of new kinds of Ni--Co-base superalloys}},
  year = {2009},
  journal = {Materials Science and Engineering: A},
  volume = {510--511},
  pages = {250--255},
  doi = {10.1016/j.msea.2008.04.128},
  shortjournal = {Mater. Sci. Eng. A}
}

@techreport{Gabb2002ME3,
  author = {Gabb, Timothy P. and Telesman, Jack and Kantzos, Peter T. and O'Connor, Kenneth},
  title = {{Characterization of the Temperature Capabilities of Advanced Disk Alloy ME3}},
  year = {2002},
  institution = {National Aeronautics and Space Administration},
  number = {NASA/TM-2002-211796},
  url = {https://ntrs.nasa.gov/citations/20020081280},
}

@article{Schulz2023IncipientMelting,
  author = {Schulz, B. and Leitner, T. and Primig, S.},
  title = {In-situ observation of the incipient melting of borides and its effect on the hot-workability of {Ni}-based superalloys},
  journal = {Journal of Alloys and Compounds},
  year = {2023},
  volume = {956},
  pages = {170324},
  doi = {10.1016/j.jallcom.2023.170324},
  shortjournal = {J. Alloys Compd.}
}

@article{Chen2023HotWorkability,
  author = {Chen, Tso-Wei and Wang, Pei-Te and Kang, Yung-Chang and Wu, Bo-Cheng and Tsou, Nien-Ti and Glatzel, Uwe and Gorsse, St{\'e}phane and Yeh, An-Chou},
  title = {Grain boundary serration tuning and its effect on hot workability of a wrought superalloy},
  journal = {Journal of Alloys and Compounds},
  year = {2023},
  volume = {960},
  pages = {170620},
  doi = {10.1016/j.jallcom.2023.170620},
  shortjournal = {J. Alloys Compd.}
}

@article{Chen2015Cooling,
  author = {Chen, Y. Q. and Francis, E. and Robson, J. and Preuss, M. and Haigh, S. J.},
  title = {Compositional variations for small-scale gamma prime ({$\gamma'$}) precipitates formed at different cooling rates in an advanced {Ni}-based superalloy},
  journal = {Acta Materialia},
  year = {2015},
  volume = {85},
  pages = {199--206},
  doi = {10.1016/j.actamat.2014.11.009},
  shortjournal = {Acta Mater.}
}

@inproceedings{Doi1984Shape,
  author = {Doi, M. and Miyazaki, T.},
  title = {The Effect of Elastic Interaction Energy on the Shape of {$\gamma'$}-Precipitate in {Ni}-Based Alloys},
  booktitle = {Superalloys 1984},
  year = {1984},
  pages = {543--552},
  publisher = {The Metallurgical Society of AIME},
  doi = {10.7449/1984/Superalloys_1984_543_552}
}

@article{Hong2009Serration,
  author = {Hong, H. U. and Kim, I. S. and Choi, B. G. and Kim, M. Y. and Jo, C. Y.},
  title = {The effect of grain boundary serration on creep resistance in a wrought nickel-based superalloy},
  journal = {Materials Science and Engineering: A},
  volume = {517},
  number = {1--2},
  pages = {125--131},
  year = {2009},
  doi = {10.1016/j.msea.2009.03.071},
  shortjournal = {Mater. Sci. Eng. A}
}

@article{Li2022NiCrOxidation,
  author = {Li, Longjun and Wang, Li and Liang, Zhida and He, Junyang and Qiu, Jingwen and Pyczak, Florian and Song, Min},
  title = {Effects of {Ni} and {Cr} on the high-temperature oxidation behavior and mechanisms of {Co}- and {CoNi}-base superalloys},
  journal = {Materials \& Design},
  year = {2022},
  volume = {224},
  pages = {111291},
  doi = {10.1016/j.matdes.2022.111291},
  shortjournal = {Mater. Des.}
}

@techreport{SpecialMetals2004Waspaloy,
  author      = {{Special Metals Corporation}},
  title       = {Waspaloy},
  institution = {Special Metals Corporation},
  number      = {SMC-011},
  year        = {2004},
  month       = sep,
  url         = {https://www.specialmetals.com/documents/technical-bulletins/waspoloy.pdf}
}

@inproceedings{Heck2020Waspaloy,
  author    = {Heck, Karl A. and Zhou, Ning and Kernion, Samuel J.
               and Rickert, Danielle and {Van Weereld}, Filip},
  title     = {A New Co-free Ni-Based Alloy for Gas Turbine
               and Exhaust Valve Applications},
  booktitle = {Superalloys 2020},
  publisher = {Springer},
  address   = {Cham},
  year      = {2020},
  pages     = {142--152},
  doi       = {10.1007/978-3-030-51834-9_14}
}
}
\end{document}